\pdfoutput=1

\documentclass[fleqn,usenatbib,letters]{mnras}

\usepackage{newtxtext,newtxmath}

\usepackage[T1]{fontenc}
\usepackage{ae,aecompl}

\usepackage{graphicx}	
\usepackage{amsmath}	
\usepackage{amssymb}	
\usepackage{bm}
\usepackage{tabularx, booktabs}

 \hypersetup{pdfauthor={T. Tröster},
               	     pdftitle={Painting with baryons: augmenting N-body simulations with gas using deep generative models},
                      bookmarksnumbered=true}
               
\newcommand{\paper}{\textit{Letter}}
\newcommand{\SLICS}{\mbox{SLICS}}
\newcommand{\BAHAMAS}{\mbox{BAHAMAS}}
\newcommand{\WMAP}{\mbox{WMAP9}}

\newcommand{\units}[1]{\ensuremath{\,\mathrm{#1}}}
\newcommand{\hMpc}{\ensuremath{\units{h^{-1}Mpc}}}
\newcommand{\degree}{\ensuremath{^\circ}}
\newcommand{\abs}[1]{\left|#1\right|}

\renewcommand{\vec}{\bm}
\newcommand{\mat}{\mathbf}

\newcommand{\diff}{\mathrm{d}}

\newcommand{\E}[2][]{\mathbb{E}_{#1}\left[#2\right]}

\newcommand{\id}{\mat{I}}

\newcommand{\normal}{\mathcal{N}}

\newcommand{\DKL}[2]{\mathbb{D}_\mathrm{KL}\left(#1||#2\right)}

\newcommand{\approptoinn}[2]{\mathrel{\vcenter{
  \offinterlineskip\halign{\hfil$##$\cr
    #1\propto\cr\noalign{\kern2pt}#1\sim\cr\noalign{\kern-2pt}}}}}

\title[Painting baryons]{Painting with baryons: augmenting $N$-body simulations with gas using deep generative models}

\author[T. Tr\"oster et al.]{Tilman Tr\"oster,$^{1}$\thanks{E-mail: ttr@roe.ac.uk}
Cameron Ferguson,$^{1}$
Joachim Harnois-D\'eraps,$^{1}$
\newauthor{and Ian~G.~McCarthy$^{2}$}
\\
$^{1}$Institute for Astronomy, University of Edinburgh, Royal Observatory, Blackford Hill, Edinburgh, EH9 3HJ, UK\\
$^{2}$Astrophysics Research Institute, Liverpool John Moores University, 146 Brownlow Hill, Liverpool L3 5RF\\
}

\date{Accepted XXX. Received YYY; in original form ZZZ}

\pubyear{2019}

\begin{document}
\label{firstpage}
\pagerange{\pageref{firstpage}--\pageref{lastpage}}
\maketitle

\begin{abstract}
Running hydrodynamical simulations to produce mock data of large-scale structure and baryonic probes, such as the thermal Sunyaev-Zeldovich (tSZ) effect, at cosmological scales is computationally challenging.
We propose to leverage the expressive power of deep generative models to find an effective description of the large-scale gas distribution and temperature. 
We train two deep generative models, a variational auto-encoder and a generative adversarial network, on pairs of matter density and pressure slices from the BAHAMAS hydrodynamical simulation.
The trained models are able to successfully map matter density to the corresponding gas pressure.
We then apply the trained models on 100 lines-of-sight from SLICS, a suite of $N$-body simulations optimised for weak lensing covariance estimation, to generate maps of the tSZ effect.
The generated tSZ maps are found to be statistically consistent with those from BAHAMAS.  
We conclude by considering a specific observable, the angular cross-power spectrum between the weak lensing convergence and the tSZ effect and its variance, where we find excellent agreement between the predictions from BAHAMAS and SLICS, thus enabling the use of SLICS for tSZ covariance estimation.
\end{abstract}

\begin{keywords}
large-scale structure of Universe -- intracluster medium -- methods: numerical
\end{keywords}


\section{Introduction}
One of the main challenges in the full exploitation of current and future weak lensing datasets is our limited understanding of the effect of baryonic processes, such as feedback from active galactic nuclei (AGN), on the distribution of matter in the Universe. 
Hydrodynamical simulations are in principle capable of providing a full description of the distribution of all matter components but predictions of the clustering of  matter  currently differ significantly between simulation codes \citep{Chisari2018}. 
This uncertainty in the modelling of the matter distribution will lead to significant biases in the cosmological parameter inference if not accounted for \citep{Huang2018}.
It is thus important to identify observations that can be used to calibrate these simulations and reject those that are unable to reproduce observations.

One such observable is the thermal Sunyaev-Zeldovich (tSZ) effect \citep{Sunyaev1972}, a measure of the electron pressure in the Universe. 
Its linear dependence on the gas density and redshift-independence makes the tSZ effect an attractive observable to characterise the distribution of baryons and the processes that affect it. 
The availability of tSZ maps covering the full sky have 
spurred a wide range of analyses that cross-correlate the tSZ effect with probes of large-scale structure \citep[see, e.g., ][for a review]{Mroczkowski2018}.

A major challenge in these analyses of tSZ data is the estimation of the covariance matrix. 
The computational complexity of hydrodynamical simulations makes it prohibitively expensive to run them in the necessary numbers and volumes necessary for robust estimates of the covariance.
Much of the computational effort of cosmological hydrodynamical simulations goes into simulating processes that couple very small scales, such as the accretion of gas onto super-massive black holes, to very large scales, like outflows of hot gas into the intracluster medium (ICM).
Current large-area tSZ data lacks the resolution to resolve these small-scale processes, however.
It is thus conceivable that there exists an effective description of the large-scale distribution of gas and its properties that can be computed more efficiently than running a full cosmological hydrodynamical simulation. 
If we can find such an effective description that only depends on the dark matter distribution, we can use it to augment existing $N$-body simulations with gas and create mock tSZ observations.

In this \paper, we consider the specific task of generating maps of the tSZ effect for the Scinet Light Cone Simulations \citep[SLICS,][]{Harnois-Deraps2018}, a suite of $N$-body simulations designed for weak lensing covariance estimation. 
This allows us to leverage the close to thousand independent SLICS realisations to estimate the covariance matrix of the cross-correlations between the tSZ effect and large-scale structure probes.
We show as a proof-of-concept that a class of machine learning methods -- deep generative models -- can map the (dark) matter SLICS density to pressure, thus allowing us to create tSZ maps for SLICS.

Deep generative models allow the creation of synthetic data whose statistical properties match those of some training data set.
Their power and versatility have already seen them being adapted for astrophysical applications, for example in the generation of galaxy images \citep{Ravanbakhsh2016} and different tracers of large-scale structure \citep{Rodriguez2018,He2018,Zhang2019,Kodi-Ramanah2019}.
In this work, we consider two classes of deep generative models: variational auto-encoders (VAE) and generative adversarial networks (GAN). 
Variational auto-encoders \citep{Kingma2013,Rezende2014} and their conditional formulations \citep[e.g.,][]{Sohn2015} express the problem as a graphical model, where the distributions are usually modelled using neural networks.
Their clear probabilistic interpretation and stable training behaviour makes VAE an attractive choice for generative models.
Generative adversarial networks \citep{Goodfellow2014} have been used to achieve many of the recent state-of-the-art results in deep generative modelling \citep[e.g.,][]{Karras2018b}. 
While GAN tend to outperform VAE in the quality of their outputs, achieving stable training is considerably more challenging.

\section{Data}
We make use of two simulation suites in this work: SLICS and BAHAMAS \citep{McCarthy2017}, a suite of calibrated hydrodynamical simulations for large-scale structure cosmology. 
We wish to augment the (dark) matter-only SLICS with baryons in order to leverage the large number of independent volumes for tSZ covariance estimation. 
To this end, we use BAHAMAS to create a training set of matching pairs of matter density and pressure slices, which allows us to train our deep generative models to predict the tSZ effect for SLICS.

\subsection{SLICS}
SLICS are a suite of $N$-body simulations consisting of 932 independent $(505\hMpc)^3$ volumes, designed for weak lensing covariance estimation. 
Due to the large number of realisations, no particle snapshot are kept. 
Instead, at 18 redshifts between $z=0$ and $z=3$, half of the volume (252.5{\hMpc}) is projected into a two-dimensional mass plane. 
This lack of three-dimensional particle snapshots is not a restriction for our application, since the tSZ effect is, like lensing, a projected quantity.

Beside the mass planes, which are of constant comoving size, SLICS also provide light-cone density maps -- hereafter called `delta maps` -- which have a constant angular size of 10 degree, and 
lensing convergence maps matched to the distribution of source galaxies of contemporary weak lensing surveys.
In this work we use the convergence maps corresponding to a single wide $z\in (0.1, 0.9)$ redshift bin of the Kilo-Degree Survey \citep{Hildebrandt2017}.

\subsection{BAHAMAS}
\label{sec:training-data}
BAHAMAS is a suite of hydrodynamical simulations that implements stellar and active galactic nuclei (AGN) feedback, tuned to recover the present-day galaxy stellar mass function and baryon fractions of massive systems.
While the suite includes runs with massive neutrinos and changes to the AGN feedback \citep{McCarthy2018}, we restrict ourselves to the case of `TUNED' AGN feedback and no massive neutrinos. 
For this case three independent volumes exists, with particle snapshots at 15 redshifts between $z=0$ and $z=3$. 
Beside the particle data for three volumes, BAHAMAS also provides lensing convergence maps (same $z\in(0.1, 0.9)$ KiDS n(z) as used for SLICS) and maps of the tSZ effect for 25 line-of-sight (LOS).

In order to train deep generative models that map the SLICS matter density to pressure maps, we need to create matter density planes from BAHAMAS that match the statistical properties of those from SLICS. 
Both SLICS and BAHAMAS use \WMAP-based cosmologies, albeit slightly different ones. 
In this work we neglect this small difference between the cosmologies, such that the problem reduces to creating 252.5{\hMpc} thick matter slices from the BAHAMAS boxes. 
Taking such slices out of the BAHAMAS volumes would only yield a small number of matter density planes, while training deep learning models often requires thousands of training samples. 
To reach the required number of training samples, we instead create slices of thickness 100{\hMpc} and 150{\hMpc}, such that we can form combinations that have an effective thickness of 250{\hMpc}, close to that of the SLICS matter density planes.
Using the three BAHAMAS volumes and three projection directions, we are able to create 14 such slices, which we furthermore split into $4^2$ tiles, with $512^2$ pixels each. 
This splitting into tiles has two advantages: it increases the number of possible combinations 256-fold and it allows for sufficiently small pixels (0.2\hMpc), while not exceeding memory limits during training.

For the training set we use 11 of the 14 slices, allowing us to form $30\,976$ samples per redshift.
The remaining 3 slices are used as the test set, which comprises 2304 samples.
Finally, we rescale the dark matter-only density planes by $\frac{\Omega_\mathrm{m}}{\Omega_\mathrm{c}}$, i.e., the ratio of total matter to dark matter, accounting for the fact that SLICS assumes that all matter is dark matter.
This ignores the effect of baryonic processes on the dark matter distribution. 
Our main objective is the generation of tSZ maps, however, for which this back-reaction effect on dark matter is negligible. 

Using the gas density and temperature, we produce pressure tiles analogously to the process for the matter tiles outlined above. 
These pairs of matter density and pressure tiles form the training and test sets for the remainder of this work.

\section{Methods}

\begin{figure*}
	\begin{center}
		\includegraphics[width=0.32\textwidth]{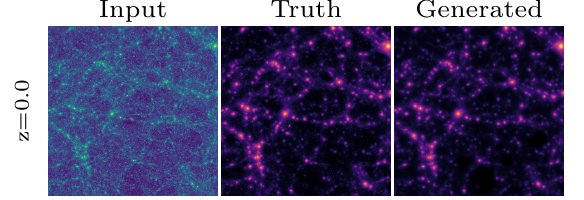}
		\includegraphics[width=0.32\textwidth]{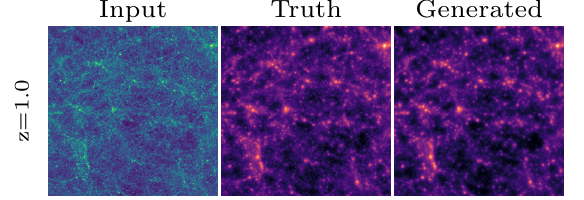}
		\includegraphics[width=0.32\textwidth]{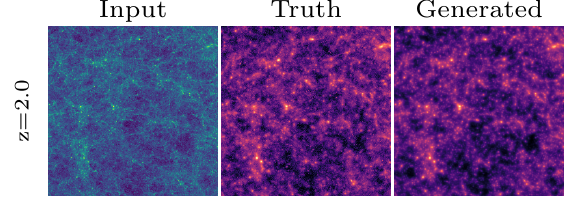}
		\vspace{-1mm}
		\caption{Comparison of a random input dark matter tile to the ground truth pressure tile from BAHAMAS and the pressure tile generated by the VAE.}\vspace{-5mm}
		\label{fig:samples}
	\end{center}
\end{figure*}

\subsection{Variational auto-encoders}
\label{sec:vae}
The basic quantity we want to find is the distribution $p(\vec x|\vec y)$ of the pressure field $\vec x$ given the (dark) matter density field $\vec y$. 
The conditional probability $p(\vec x|\vec y)$ describes processes that usually require full hydrodynamical simulations to model. 
To capture this rich mapping between dark matter and pressure, we introduce a latent variable $\vec z$ that describes some internal representation of this mapping. 
The conditional probability $p(\vec x|\vec y)$ can now be written as \citep{Sohn2015}
\begin{equation}
	p(\vec x|\vec y) = \int \diff\vec z p(\vec x, \vec z|\vec y) =  \int \diff\vec z p(\vec x|\vec y,\vec z)p(\vec z|\vec y) \ .
\end{equation}
This can be seen as an infinite mixture model, where the mixture is made up of different models $p(\vec x|\vec y,\vec z)$ of the mapping between dark matter and pressure, parametrised by $\vec z$, and weighted by the prior $p(\vec z|\vec y)$, which describes the dependence of $\vec z$ on the dark matter field $\vec y$.
It is possible to derive a lower bound on the log-probability $\log p(\vec x | \vec y)$, the evidence lower bound (ELBO) \citep[see, e.g.,][]{Sohn2015}:
\begin{multline}
	\label{equ:elbo}
    \log p(\vec x | \vec y) \ge -\DKL{q_\phi(\vec z |\vec x, \vec y)}{p_{\theta_1}(\vec z|\vec y)} \\
    + \E[\vec z\sim q_\phi(\vec z|\vec x, \vec y)]{\log p_{\theta_2}(\vec x | \vec y, \vec z)} \ ,
\end{multline}
where $\DKL{\cdot}{\cdot}$ denotes the Kullback-Leibler divergence and $\E[z\sim p]{f}$ the expectation of $f$ with respect to $p(z)$.
The first term describes the difference between $q_\phi(\vec z | \vec x, \vec y)$, an approximation to $p(\vec z|\vec x,\vec y)$ called the recognition network, and the prior $p_{\theta_1}(\vec z|\vec y)$. 
The second term captures the performance of the generator network $p_{\theta_2}(\vec x | \vec y, \vec z)$, expressed as the expectation of the log-likelihood of the pressure field $\vec x$, conditioned on the dark matter field $\vec y$ and latent variable $\vec z$.

The recognition network $q_\phi(\vec z | \vec x, \vec y)$, prior $p_{\theta_1}(\vec z|\vec y)$, and generator network $p_{\theta_2}(\vec x | \vec y, \vec z)$ are all modelled as multivariate Gaussian distributions where the mean and variance are predicted by convolutional neural networks.
The parameters $\phi$, $\theta_1$, and $\theta_2$ of these networks can be efficiently optimised using stochastic gradient ascent.
The details of the implementation are described in Appendix~\ref{sec:implementation}.
In order to sample the pressure field given a dark matter density field $\vec y$, we first sample $\vec z$ from the prior $\vec z \sim p_{\theta_1}(\cdot |\vec y)$ and then sample the pressure field $\vec x$ from the generator $\vec x \sim p_{\theta_2}(\cdot | \vec y, \vec z)$.

The second term in Eq.~\eqref{equ:elbo} quickly becomes dominated by the determinant of the predicted variance, at which point most of the training time is spent optimising the prediction of the variance without improving the prediction of the mean. 
If we choose to use the predicted mean as the output instead of sampling from $p_{\theta_2}(\vec x | \vec y, \vec z)$, estimating the variance is not required and we can speed up the training significantly by assuming a fixed variance.
For our fiducial model, we follow this approach and fix the variance to the identity, i.e., $p_{\theta_2}(\cdot | \vec y, \vec z) = \normal(\vec\mu_{\theta_2}(\vec y, \vec z), \id)$. 
We have verified that allowing the model to vary the variance yields consistent results, albeit at the cost of significantly slower convergence.

\subsection{Generative adversarial networks}
Generative adversarial networks cast the generative process into a game between a generator $G : (\vec y \mapsto \vec x_G)$, that maps dark matter density $\vec y$ to pressure map $\vec x_G$ and a discriminator $D : (\vec x, \vec y\mapsto [0,1])$, that tries to determine whether a given sample has come from the training set or has been generated by $G$, assigning 1 in the former case and 0 in the latter.
Unlike in the case of the VAE, where the performance of the generator is quantified by the $\chi^2$ statistic of the pixel values (or $L_2$-norm in case of a fixed variance), the GAN is able to learn an optimal discriminator, thus allowing for more expressive power. 

Both $G$ and $D$ are represented by convolutional neural networks, parametrised by $\theta_{G}$ and $\theta_{D}$. 
Training proceeds by minimising the cost functions $J^{(D)}$ and $J^{(G)}$, where the cost function of the discriminator is
\begin{equation}
	\label{equ:discr-loss}
	J^{(D)} = -\frac{1}{2}\E{\log D(\vec{x},\vec{y})} - \frac{1}{2}\E{\log(1 - D(G(\vec{y}),\vec{y}))} \ .
\end{equation}
Both expectations are with respect to the distribution of the data, i.e., the pressure map $\vec{x}$ and dark matter density $\vec{y}$ are drawn from the training set. 

In a zero-sum game, the generator's cost function would be the same as that of the discriminator but with opposing sign. 
Here we instead take the approach of using the `heuristic non-saturating loss' proposed by \cite{Goodfellow2016}:
\begin{equation}
	\label{eq:equ:gen-loss}
	J^{(G)} = -\frac{1}{2}\E{\log(D(G(\vec{y}),\vec{y}))} + \mathcal{L}_{\mathrm{perceptual}} \ ,
\end{equation}
where the perceptual term $\mathcal{L}_{\mathrm{perceptual}}$ is given by
\begin{equation}
	\label{equ:perceptual-loss}
	\mathcal{L}_{\mathrm{perceptual}} = \lambda_{\mathrm{perceptual}}\  \E{\abs{\vec{x} - G(\vec{y})}} \ .
\end{equation}
The perceptual term therefore captures the difference (under the $L_1$-norm) between the true pressure map $\vec{x}$ and the sample which $G$ produces.
The parameter $\lambda_{\mathrm{perceptual}}$ controls the relative weighting between the generator's adversarial and perceptual loss and is chosen such that those two losses have the same order of magnitude in the early stages of training. 

\subsection{Light-cone generation}
\label{sec:light-cone}
To create the light-cones, we first produce pressure maps corresponding to the SLICS delta maps. 
For the two lowest redshift slices ($\bar z=0.042$ and $\bar z=0.130$), the $(100\hMpc)^2$ tiles used by the deep generative models are larger than the light-cone. 
For these redshifts we take $(100\hMpc)^2$ cutouts from the full SLICS mass planes, centred on the light-cone, and let the generative models predict the corresponding pressure maps. 
The maps are then cropped to the size of the light-cone.

For higher redshifts, the light-cone is larger than the physical size of the tiles. We choose $(100\hMpc)^2$ cutouts from the SLICS delta maps, ensuring at least 20\% overlap between adjacent cutouts. 
These cutouts are then fed through the generative models. 
In the overlap regions, the predicted pressure maps are averaged, taking into account a weighting scheme that down-weights pixels at the tile border to minimise edge effects.

This tiling process in principle preserves modes larger than the tile size since the generative models are sensitive to the mean of the input tile. 
Physical correlations beyond the dependence on the mean on scales of $\approx 100\hMpc$ are not captured, however. 
Since those scale are in the linear regime, this only incurs a negligible bias.

Finally, the pressure maps are converted into tSZ maps, following the prescription in \citet{McCarthy2018}. We use \texttt{CCL} \citep{Chisari2019} for the computation of cosmological background quantities.

\section{Results}

\begin{figure}
	\begin{center}
		\includegraphics[width=\columnwidth]{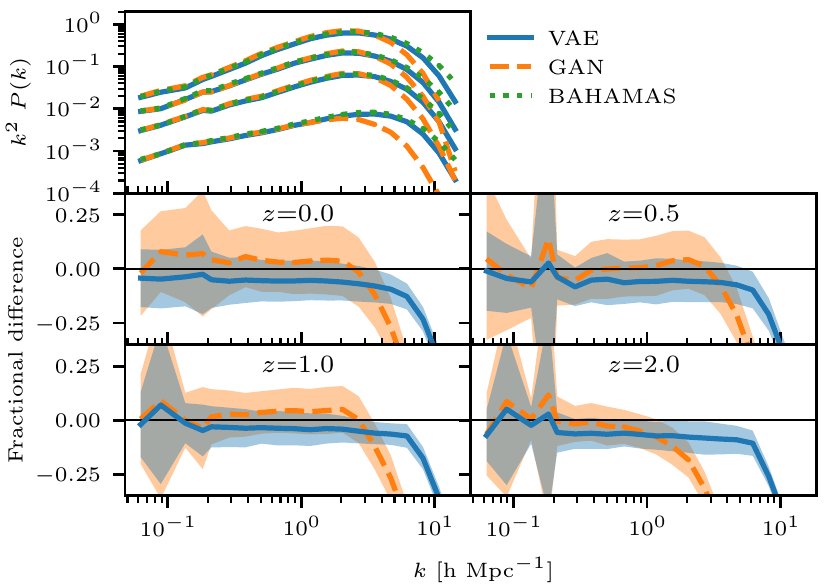}
		\vspace{-4mm}\caption{\emph{Top panel}: cross-power spectra between dark matter tiles from the test set and pressure tiles created by the VAE (blue), GAN (orange), and the truth from BAHAMAS (green) for redshifts 0.0, 0.5, 1.0, and 2.0. \emph{Bottom panels}: fractional difference between the true dark matter--pressure cross-spectra from BAHAMAS and those predicted by the VAE (blue) and GAN (orange). The shaded region denotes the one-standard deviation range.}\vspace{-3mm}
		\label{fig:frac-diff-tile}
	\end{center}
\end{figure}

In order for the generative models to be able to predict tSZ maps for SLICS, they need to meet two requirements: firstly, the cross-spectrum between matter and pressure on the tiles need to match on the test set. 
Secondly, the models need to be able to interpolate between redshifts, since the SLICS mass plane redshifts do not match those of the BAHAMAS snapshots. 
We discuss the performance on these two requirements in the next two subsections before considering the full case of the cross-correlation between lensing and the tSZ effect.

\subsection{Performance on tiles}
\label{sec:tile-performance}
In Fig.~\ref{fig:samples} we show a random triplet of input dark matter, true pressure, and generated pressure tiles for three redshifts, while Fig.~\ref{fig:frac-diff-tile} shows the cross-power spectrum between matter and pressure tiles for four redshifts between $z=0$ and $z=2$.
Both the VAE and the GAN are able to reproduce the cross-power spectrum up to $k\lessapprox 2$, with the agreement generally being within 20\%. 
These deviations are negligible compared to the intrinsic scatter of the signal between tiles. 
Since each redshift slice is made up of multiple tiles, and multiple redshift slices contribute to the final tSZ map, the deviations further average out. 
We thus conclude that the models are able to reproduce the cross-spectrum to a degree sufficient for our application.
The VAE shows a constant offset for the cross-power spectrum, while the GAN has a poorer performance at small scales, neither of which would average out. 
Further optimisation of the hyper parameters of the models and its architecture would likely be able to ameliorate this. 
Since the effect is small, we leave these optimisations for future work.

Our objective is the generation of mock tSZ data.
The low resolution of current tSZ washes out small-scale information, such that the deficit of our models at small scales can be safely ignored.

\subsection{Redshift interpolation}
To test whether the models are capable of interpolating between redshifts, we train them on all redshifts except $z=0.25$ and then validate the models at this redshift. 
Both the VAE and the GAN are able to predict the $z=0.25$ cross-spectrum as well as when that redshift is included in the training set.
This is likely due to the fact that the redshift evolution is a continuous process that is well constrained by the remaining 10 redshift slices.

\subsection{Performance on light-cones}
\begin{figure}
	\begin{center}
		\includegraphics[width=\columnwidth]{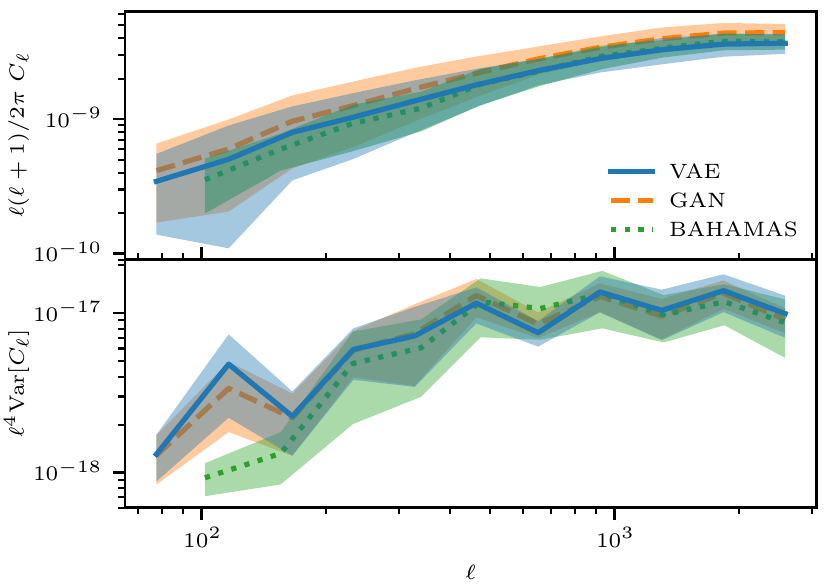}
		\vspace{-4mm}\caption{\emph{Top panel}: angular cross-power spectra between the lensing convergence and the tSZ effect. The cross-power spectrum predicted by {\BAHAMAS} is shown in green and is assumed to be the truth. The cross-power spectra between the SLICS convergence and the tSZ map generated with the VAE (GAN) are shown in blue (orange). The shaded region denotes the one-standard deviation range.
		\emph{Bottom panel}: variance of angular cross-power spectra between the lensing convergence and the tSZ effect, estimated from 25 {\BAHAMAS} and 100 {\SLICS} LOS. The variance estimates are corrected for differences between {\BAHAMAS} and {\SLICS} due to cosmic variance. The errors on the variance are estimated by bootstrapping.}\vspace{-3mm}
		\label{fig:tSZ-spectra}
	\end{center}
\end{figure}
We run both the VAE and GAN on 100 SLICS light-cones and compute the cross-spectrum between the lensing convergence and generated tSZ map.
We then compare this to the same cross-spectrum computed from 25 BAHAMAS LOS.
Fig.~\ref{fig:tSZ-spectra} comprises the main results of this work, demonstrating the agreement between {\BAHAMAS} and {\SLICS} on the predicted lensing-tSZ cross-correlation and its variance. 
The agreement is very good and well within the intrinsic scatter due to sample variance.
The VAE recovers the cross-spectrum to percent-level at small scales and to within 10\% at large scales.
The GAN slightly over-predicts the cross-spectrum by around 20\%, which is consistent with the different performances of the two models on the tiles, see Sec.~\ref{sec:tile-performance} and Fig.~\ref{fig:frac-diff-tile}.

The LOS used in this comparison are different for BAHAMAS and SLICS and the variance estimates are therefore affected by cosmic variance.
To estimate the effect of cosmic variance, we approximate tSZ maps as being proportional to the convergence map by some scale-dependent factor $\alpha_\ell = {C^{\kappa y}_\ell}/{C^{\kappa \kappa}_\ell}$.
Under this simplistic approximation, the variance of the shear-tSZ cross-power spectrum is \mbox{$\mathrm{Var}[C^{\kappa y}_\ell]\approx \alpha_\ell^2 \mathrm{Var}[C^{\kappa \kappa}_\ell]$}. 
We then use the measured shear-shear variances $\mathrm{Var}[C^{\kappa \kappa}_\ell]$ from BAHAMAS and SLICS to rescale the measured SLICS shear-tSZ variance.
Even after this rescaling to account for cosmic variance, our set of SLICS light-cones overestimate the signal and variance at large scales.
These scales are dominated by the first redshift slice, which is strongly affected by differences in the light-cone generation between SLICS and BAHAMAS and which is, due to its small tangential size (21.8\hMpc), particularly strongly affected by sampling variance. 

While the objective of this work is to create tSZ maps for cross-correlation studies, we find that the auto-correlations of the tSZ maps also match well to those from BAHAMAS. 
The techniques presented in this {\paper} can therefore also be applied to estimate the covariance of the tSZ auto-correlation, such as the power spectrum.
The larger angular size of the SLICS light-cone compared to {\BAHAMAS} (10{\degree} vs 5\degree), also allows us to extend to prediction of the cross-spectra to larger scales, which is crucial for the analysis of current wide-field surveys.

\vspace{-3mm}
\section{Conclusions}
In this {\paper}, we have provided a proof-of-concept of using deep generative models to augment existing $N$-body simulations with baryons.
The generative models were trained to generate the gas pressure distribution from only on the (dark) matter density, and were subsequently employed to generated maps of the tSZ effect for SLICS, a suite of existing $N$-body simulations.
We showcased the performance of our generative models in reproducing, to a remarkable extent, the summary statistics of interest, namely the angular (cross-) power spectra, from BAHAMAS, a full hydrodynamical simulation.
Once trained, these models allow rapid generation of tSZ mock data: on the order of one CPU-hour per light-cone, compared to the $\mathcal{O}(10^5)$ CPU-hours required for a run of BAHAMAS.

In this work we restricted ourselves to the particle information in the form of matter density and pressure maps from BAHAMAS, both for training and validation. 
It will be fruitful to compare our models against other, more physically motivated models for the same observational quantities, as well as testing the models on other statistics, such as the tSZ one-point function or pressure profiles. 
We leave these detailed characterisations to future work. 

We have only scratched the surface of the potential of generative models in this work.
Possible future avenues are for example the use of different representations of the training data, such as halo catalogues or the raw particle data.
The models can also be extended to predict other quantities, such as the distribution of galaxies and X-rays emissions. 
This would allow the creation of extremely rich mock catalogues for $N$-body simulations, from large-scale structure observables, like weak lensing and galaxy clustering, to baryonic probes, such as various SZ effects and X-rays.

\vspace{-3mm}
\section*{Acknowledgements}
We thank Alexander Mead for providing the code to process the {\BAHAMAS} particle data, Fran\c{c}ois Lanusse for useful discussions, and Eric Tittley for use of his GPGPU workstation.  
This project has received funding from the European Union's Horizon 2020 research and innovation programme: TT acknowledges support under the Marie Sk\l{}odowska-Curie grant agreement No.~797794, while JHD and IGM acknowledge support from the European Research Council under grant agreements No.~647112 and No.~769130, respectively.

\footnotesize{Computations for the SLICS $N$-body simulations were enabled in part by support provided by Compute Ontario (www.computeontario.ca), Westgrid (www.westgrid.ca) and Compute Canada (www.computecanada.ca). 
The BAHAMAS simulations used the DiRAC@Durham facility managed by the Institute for Computational Cosmology on behalf of the STFC DiRAC HPC Facility (https://dirac.ac.uk/). The equipment was funded by BEIS capital funding via STFC capital grants ST/K00042X/1, ST/P002293/1, ST/R002371/1 and ST/S002502/1, Durham University and STFC operations grant ST/R000832/1. DiRAC is part of the National e-Infrastructure.}



\vspace{-3mm}
\bibliographystyle{mnras}
\bibliography{references.bib} 



\appendix
\vspace{-3mm}
\section{Implementation details}
\label{sec:implementation}
Both the VAE and GAN take dark matter density tiles and their redshifts as the input. 
The redshifts are provided as a constant feature map, i.e., each dark matter density tile gets concatenated with a tile of the same size and with a constant value containing the redshift.
The generator networks are adapted from \citet{Johnson2016}, since the problem of style transfer is qualitatively similar to our problem of translating one tracer of large-scale structure (dark matter) to another (gas pressure).
The trained models, code, detailed network architectures, and training schedules are available at \url{https://www.github.com/tilmantroester/baryon_painter/}.

The VAE has $1.6\times 10^6$ trainable parameters, while the GAN uses $5.6\times10^6$ parameters in total. This difference is largely due to the use of 4 residual blocks in the VAE compared to the 9 in the GAN and the need for a sophisticated discriminator in the GAN.

\vspace{-3mm}
\subsection{Normalisation}
A particular challenge in the prediction of pressure maps is their high dynamic range. 
Outside of haloes there is little pressure and therefore most of the pixels in the maps have values close to zero. 
The quantity of interest, the cross-spectrum between matter and pressure, is dominated by the peaks in the map, however. 
It is therefore important to ensure the few high pressure pixels are reproduced accurately.
We choose a generalised log-transform to transform the pixel distribution closer to normal:
\begin{equation}
	\label{equ:transform}
	f(\vec d) = \frac{1}{k}\log\left( \frac{\vec d}{\sigma_{\vec d}(z)}+ 1\right) \ ,
\end{equation}
where $\sigma_{\vec d}(z)$ is the standard deviation of the pixel values at redshift $z$ and $k$ is a scale parameter. We find that $k=4$ for both matter and pressure maps yields good results.

We found that using a softplus activation function for the VAE improved the generative performance significantly. 
We hypothesise that this is due to the fact that the softplus function maps low values to zero and is linear for high values.
Since most pixels in the pressure maps have low values they make a large contribution to the generator loss term in Eq.~\eqref{equ:elbo}.
Mapping these low values to zero reduces the impact of these low pixels values on the loss function.
Since the GAN only depends weakly on an $L_1$ loss through the perceptual loss Eq.~\eqref{equ:perceptual-loss}, we use a $\tanh$ activation for the generator there.


\bsp	
\label{lastpage}
\end{document}